\documentclass[aps,pra,showpacs,amsmath,amssymb,superscriptaddress,reprint,10pt]{revtex4-1}

\usepackage{graphicx}
\usepackage{dcolumn}
\usepackage{bm}
\usepackage[latin1]{inputenc}
\usepackage{epstopdf}
\usepackage{amsmath}
\usepackage{upgreek}  
\usepackage{amssymb}
\usepackage{esint}
\usepackage{xcolor}
\usepackage{latexsym}
\usepackage{mathrsfs}
\usepackage{arydshln}
\usepackage[sans]{dsfont}
\usepackage[colorlinks=true,breaklinks=true,allcolors=blue]{hyperref}

\begin{document}

\preprint{}

\title{Coherent backscattering of inelastic photons from atoms and their mirror images}

\author{P.H. Moriya}
\affiliation{Instituto de F\'{\i}sica de S\~{a}o Carlos, Universidade de S\~{a}o Paulo, C.P. 369, 13560-970 S\~{a}o Carlos, SP, Brazil}
\author{R.F. Shiozaki}
\author{R. Celistrino Teixeira}
\affiliation{Instituto de F\'{\i}sica de S\~{a}o Carlos, Universidade de S\~{a}o Paulo, C.P. 369, 13560-970 S\~{a}o Carlos, SP, Brazil}
\affiliation{Departamento de F\'{\i}sica, Universidade Federal de S\~{a}o Carlos, Rod. Washington Lu\'{\i}s, km 235 - SP-310, 13565-905 S\~{a}o Carlos, SP, Brazil}
\author{C.E. M\'aximo}
\affiliation{Instituto de F\'{\i}sica de S\~{a}o Carlos, Universidade de S\~{a}o Paulo, C.P. 369, 13560-970 S\~{a}o Carlos, SP, Brazil}
\author{N. Piovella}
\affiliation{Dipartimento di Fisica, Universit\`{a} degli Studi di Milano, Via Celoria 16, I-20133 Milano, Italy}
\author{R. Bachelard}
\affiliation{Instituto de F\'{\i}sica de S\~{a}o Carlos, Universidade de S\~{a}o Paulo, C.P. 369, 13560-970 S\~{a}o Carlos, SP, Brazil}
\author{R. Kaiser}
\affiliation{Institut Non Lin\'eaire de Nice, UMR 6618 CNRS, 1361 route des Lucioles, 06560 Valbonne, France}
\author{Ph.W. Courteille}
\email{philippe.courteille@ifsc.usp.br}
\affiliation{Instituto de F\'{\i}sica de S\~{a}o Carlos, Universidade de S\~{a}o Paulo, C.P. 369, 13560-970 S\~{a}o Carlos, SP, Brazil}

\begin{abstract}
Coherent backscattering is a coherence effect in the propagation of waves through disordered media involving two or more scattering events. Here, we report on the observation of coherent backscattering from individual atoms and their mirror images. This system displays two important advantages: First, the effect can be observed at low optical densities, which allows to work in very dilute clouds or far from resonance. Second, due to the fact that the radiation of an atom interferes constructively with that of its own image, the phenomenon is much more robust to dephasing induced by strong saturation. In particular, the contribution of inelastically scattered photons to the interference process is demonstrated.
\end{abstract}

\pacs{42.25.Fx, 32.80.Pj}
\maketitle

\section{Introduction}

Light propagating in an optically thick sample is subject to multiple scattering. Although part of the propagation can be described by a diffusion equation neglecting interferences, wave effects can alter the distribution of scattered light. In particular, disorder in the sample may lead to an enhanced scattering into the backward direction. The effect is known as coherent backscattering (CBS) in mesoscopic physics, and has been studied extensively with classical scatterers~\cite{Kuga84, Albada85, Wolf85,  Yoo1990,Mishchenko1993,Wiersma1995,Tourin1997}. The advent of laser-cooling techniques allowed to manipulate and control atomic gases, thus enabling their use as resonant and quantum scatterers. This triggered the study of coherent multiple scattering in a regime where the quantum internal structure, the wave-particle duality, and quantum statistical aspects play a role \cite{Labeyrie99,Bidel02,Kupriyanov03,Chaneliere04,Kupriyanov06}.

\begin{figure}\centering
\includegraphics[width=0.48\textwidth]{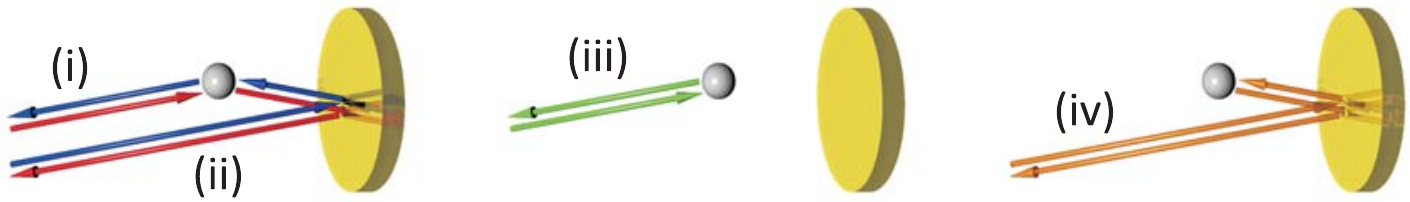}
\caption{
Four processes contributing to m-CBS. In the low intensity limit, the reciprocal paths (i) and (ii) contribute to the m-CBS fringes, whereas paths (iii) and (iv) yield a smooth background intensity.
}
\label{fig:ScatScheme}
\end{figure}

CBS is understood as resulting from the constructive interference between a scattering path involving two or more scatterers and the reciprocal (time-reversed) path (see Ref.~\cite{Wellens04}, and paths (i) and (ii) in Fig.~\ref{fig:ScatScheme}). The interference of reciprocal paths is actually robust when summed up over a large disordered sample, which was one of the surprising features in the first observation of CBS in the eighties~\cite{Kuga84,Albada85,Wolf85}. More specifically, some paths add up incoherently and result in a background radiation, whereas reciprocal paths lead to an enhanced intensity in the backward direction.
However, the quantum nature of the atoms leads to deviations in the behavior of CBS as compared to classical scatterers. For example, the presence of a Zeeman structure can break the symmetry between the two reverse paths and reduces the contrast between the enhanced peak of radiation and the background~\cite{Labeyrie99, Jonckheere00}. The time-reversal symmetry of the reciprocal multiple scattering paths is also broken in the strong driving regime, as a which-path information becomes available through the inelastically scattered photons~\cite{Bidel02,Kupriyanov06,Labeyrie08}. Such saturation-induced loss of coherence in CBS was reported with a cold strontium gas~\cite{Chaneliere04} and a cold rubidium gas ~\cite{Kupriyanov06,Labeyrie08}. Unfortunately, the theoretical treatment of saturation in multiple scattering is very challenging~\cite{KoukiTotsuka98,Wellens04,Wellens05,Balik05,Shatokhin05,Gremaud06,Wellens06,Ott13,Ketterer14} and has not been solved in full generality.

Interestingly in a different geometrical configuration coherence effects on the backscattered light can also be observed in optically thin samples, where multiple scattering is too weak to produce observable signatures. Indeed the introduction of a reflective interface -- a dielectric mirror for example -- allows for the radiation of the image scatterer to interfere constructively with that of the original scatterer, eventually resulting in a coherent backscattering process~\cite{Greffet91}. This single scattering regime here involves four processes for each atom [depicted in Fig.~\ref{fig:ScatScheme}], accounting for the real atom and its mirror image, as well as the laser and its image. Let us call $\mathbf{k}_0$ the incident wavevector, $\mathbf{k}$ the scattered wavevector and $\hat z$ the normal to the mirror. For a single atom, all four processes sum up coherently, and the resulting scattered light pattern presents full interference contrast. When the radiation of all atoms is disorder-averaged, though, all interference fringes disappear, except at wavevectors $\mathbf{k}$ such that $\mathbf{k}_0.\hat z=-\mathbf{k}.\hat z$, because at these specific directions processes (i) and (ii) have the same optical path for all atoms in the cloud. The resulting interference fringes present a circular symmetry around the mirror's normal direction, with a number of maxima depending on the spatial extension of the atomic cloud. This effect, that will henceforth be referred to as mirror-assisted Coherent BackScattering (m-CBS), has been observed for classical scattering media~\cite{Labeyrie00}.

In this work, we report on the experimental observation of m-CBS from a laser-cooled gas of strontium atoms in the presence of a dielectric mirror. A series of circular fringes predicted by the theory are experimentally observed and quantitatively analyzed. The period and envelope of the interference fringes allow, respectively, to precisely determine the position and longitudinal size of the atomic cloud. We show that in the strong field regime, where the atoms are fully saturated, the comparison between theory and experimental results allows to show that, in contrast to CBS, inelastic photons fully contribute to m-CBS.

\section {Experimental setup}

We prepare our atomic sample in a typical strontium apparatus, which we briefly describe in the following. A collimated atomic beam emerges from an array of microtubes located at the output of an oven heated to $550^\circ$C. The atomic beam is then decelerated in a $28~$cm long Zeeman Slower in spin-flip configuration by a counterpropagating laser beam of power $\sim40~$mW tuned $500~$MHz to the red of the $461~$nm resonance. The Zeeman slower beam has a 1/$e^2$ radius of $4~$mm at the entrance of the experimental apparatus, being focused onto the oven output after propagation through the whole system (about $90$cm long). The cooled strontium beam is captured in the science chamber by a magneto-optical trap (MOT) generated by three pairs of counter-propagating collimated $461~$nm laser beams and a quadrupole magnetic field; the latter is produced by a pair of coils in anti-Helmholtz configuration (axial magnetic gradient $|\nabla{B}|=70~$G/cm). Each laser beam has a $1/e^2$ radius of $5~$mm and is detuned by $-40~$MHz from resonance.  A repumping laser addressing the $497$nm $^3P_2\rightarrow{^3}D_2$ transition is used to recycle atoms that have decayed to the metastable state ${^3}P_2$. In this way, we are able to generate cold gases with $\sim10^8$ $^{88}$Sr atoms at a temperature below $10~$mK. Resonant absorption imaging reveals an approximate Gaussian density profile with a $1/\sqrt{e}$ radius of $0.9(1)~$mm.

\begin{figure}\centering
\includegraphics[width=0.48\textwidth]{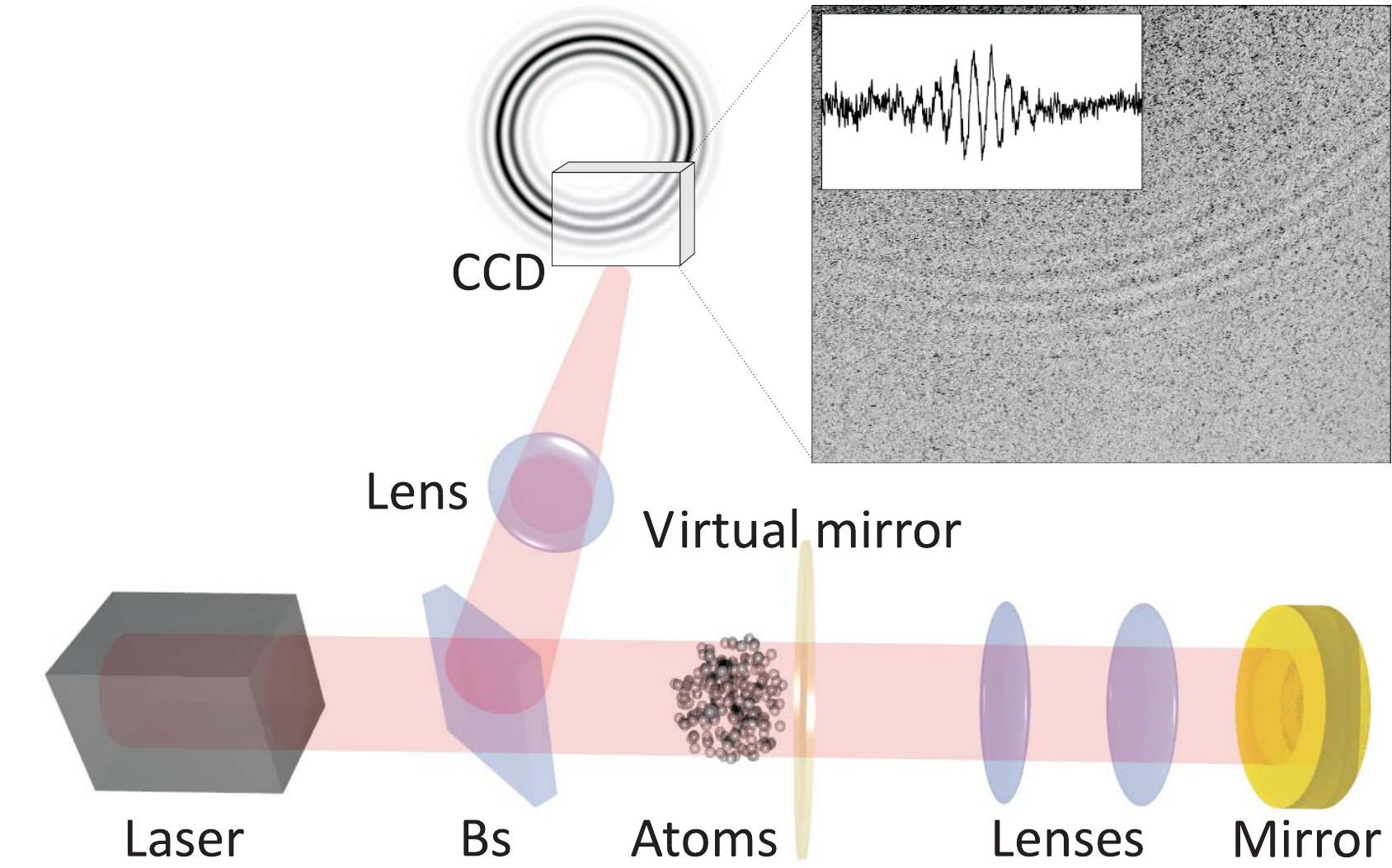}
\caption{
Experimental setup:  A laser beam passes through a beamsplitter (BS) before illuminating the atomic cloud a first time. It is then reflected on a virtual mirror (VM), created by two lenses and a physical mirror, at a small angle of $\theta_0\sim1^\circ$ with the normal of the mirror. After being reflected, it crosses the cloud again, before being sent by the beamsplitter, to a CCD camera detecting the angular distribution of the light in the focal plane of a lens. The top right picture shows an interference pattern obtained in our experimental setup, exhibiting the predicted circular symmetry. The inset is an azimuthal angular average of the picture.
}
\label{fig:ExpScheme}
\end{figure}

The setup for the m-CBS experiment is sketched in Fig.~\ref{fig:ExpScheme}. The scattering medium is a cold gas (temperature $\lesssim 10~$mK) of $^{88}$Sr atoms in its ground state ${^1}S_0$, and the transition ${^1}S_0\rightarrow{^1}P_1$ (at $\lambda=2\pi/k=461~$nm with a linewidth of $\Gamma=(2\pi)30.5~$MHz) is used for the resonant scattering. The $461~$nm probe laser beam (the m-CBS beam) has a waist of $1.5~$mm and linear polarization. It first passes through a 50-50 non-polarizing wedged beamsplitter before reaching the atoms. A combination of two lenses with focal distances $f=15~$cm and separated by a distance of $2f = 30~$cm creates a virtual image of a real mirror, placed at a distance $d$ after the last lens, at a distance $2f-d$ before the first lens. This configuration has been used to study Talbot physics and even allows negative distances to be realized ~\cite{VirtualMirror, Labeyrie14}. The m-CBS beam is reflected with an angle $\theta_0\sim1^\circ$ compared to the mirror's normal direction. Having crossed again the atomic cloud, the beam is partially reflected by the beamsplitter onto a 200mm lens. A CCD camera placed at the focal plane of the lens allows for the detection of the  angular radiation pattern. To avoid the direction of the reflected m-CBS beam, which would saturate its pixels, the CCD camera only captures part of the circular pattern. An  experimental measurement of such fluorescence pattern is shown in the top right area of Fig.~\ref{fig:ExpScheme}.

During the experimental sequence, a MOT of $10^8$ atoms with a $1/\sqrt{e}$ radius of $0.9(1)~$mm is loaded within $2~$s. After turning off the MOT cooling beams, we wait $200~\upmu$s and shine the m-CBS beam pulse of 200~$\upmu$s duration onto the atomic cloud, capturing an image in the presence of atoms. After $1$s we record a background image in the absence of atoms. We run this sequence $\approx 200$ times for the same parameters to obtain a disorder-averaged final image as shown in Fig.~\ref{fig:ExpScheme}.

\section{Linear regime}

\begin{figure}
	\includegraphics[width=7.8 truecm]{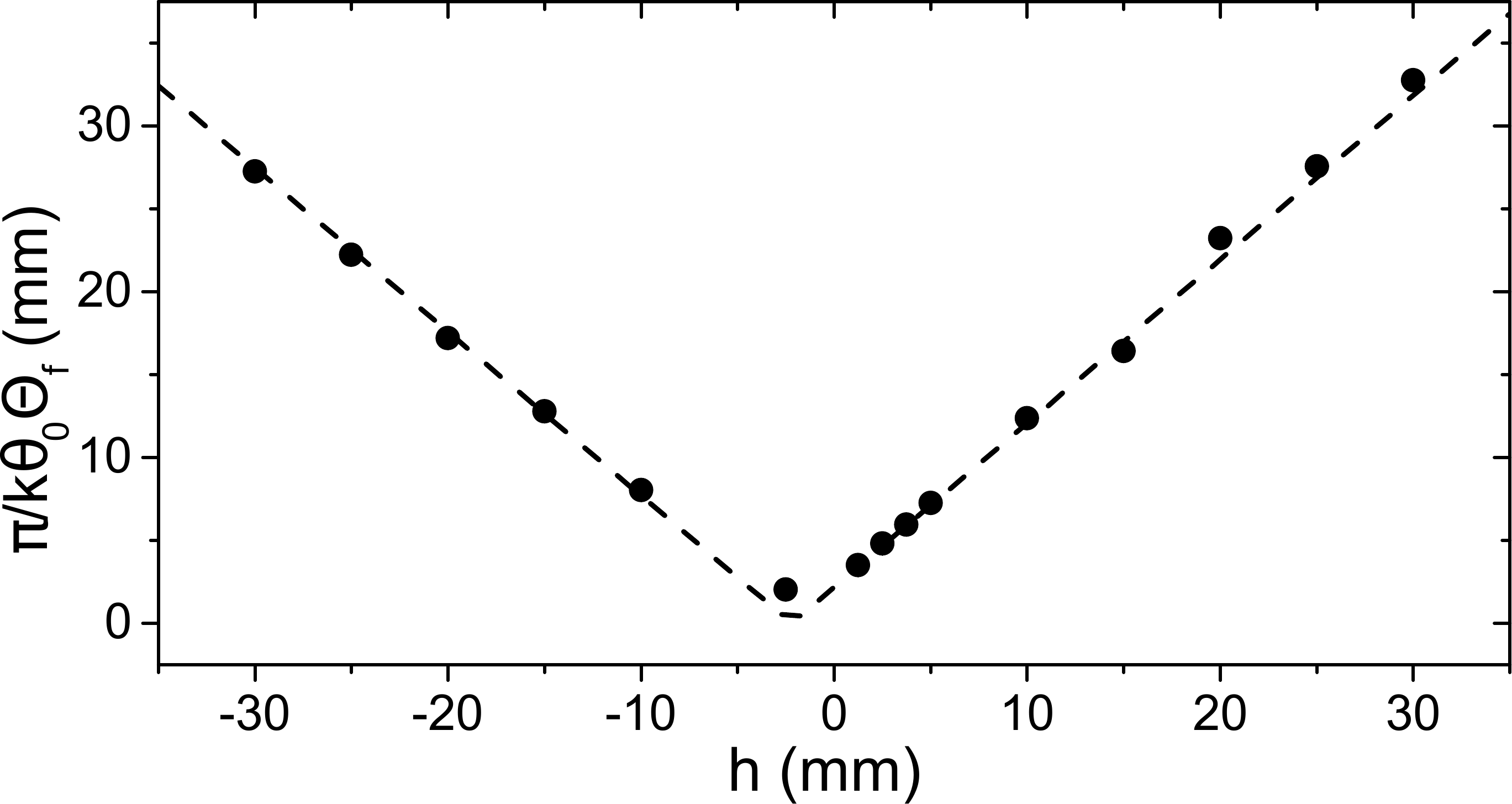}
	\caption{Mirror position expected from m-CBS theory (i.e., $h_{thr}=\pi/k\theta_0\Theta_f$) as a function of the experimental mirror position. This measurement allowed to detect an initial experimental misalignment of $x_0=2.1(1)~$mm, which corresponds to the shifted minimum of the fit $h_{thr}=|A h + x_0|$ (dashed line); we obtained $A=0.988(12)$. The error bars associated with the experimental data are smaller than the symbol size.}
	\label{fig:DistanceMirror}
\end{figure}

The interference phenomenon of m-CBS is best understood in the linear regime, i.e., for a saturation parameter $s=2\Omega_0^2/(\Delta^2+\Gamma^2/4)\ll 1$, where $\Omega_0$ is the Rabi frequency due to the incident laser and $\Delta$ is the laser detuning from the atomic resonance. In this limit the atoms behave as classical linear scatterers and the total fluorescence is a linear combination of four processes depicted in Fig.~\ref{fig:ScatScheme}. The reciprocity of processes (i) and (ii) requires to add up the corresponding field amplitudes for each atom, whereas the paths (iii) and (iv) have no reciprocal counterpart, and so the corresponding scattering intensities contribute, after disorder-averaging, to an incoherent background. In this low saturation limit ($s\ll 1$), the intensity radiated by a Gaussian cloud of atoms and its mirror image reads (see Appendix~\ref{theory})

\begin{equation}\label{eq:mod}
	I(\theta) \propto s \left[1+\frac{1}{2}e^{-2(\theta_0k\sigma_z)^2(\theta-\theta_0)^2}\cos (2\theta_0 kh(\theta-\theta_0))\right],
\end{equation}
where $h$ is the distance between the virtual mirror and the center of the atomic cloud and $\sigma_z$ is the longitudinal cloud radius at $1/\sqrt{e}$. To obtain Eq.~\eqref{eq:mod}, a small angle approximation has been applied ($\theta_0\ll 1$ and $|\theta-\theta_0|\ll\theta_0$). The second term in the bracket on the r.h.s. of  Eq.~\eqref{eq:mod} corresponds to the single scattering interference of m-CBS, surviving averaging over the atomic spatial Gaussian distribution within a Gaussian angular envelope of half-width at $1/e^2$ given by $\Phi=1/\theta_0k\sigma_z$. Since the fringes have an angular period $\Theta_f=\pi/\theta_0kh$, one typically expects to detect a number $\sim h/\pi \sigma_z$ of fringes on the scattered light. Together, both terms yield an ideal contrast of m-CBS of $C=1$, defined as $C=(I_{\text{max}}-I_{\text{min}})/I_{\text{background}}$. The fringes' period depend on the inverse of the distance of the cloud to the mirror ($\Theta_f=\pi/\theta_0 kh$), so the position of the mirror can in principle be evaluated from the fringes' pattern. To confirm this effect, the (virtual) mirror position $h$ was varied for about $6~$cm around the center of the cloud, and for each position an interference pattern similar to that of Fig.~\ref{fig:ExpScheme} was extracted. Note that we are able to place the virtual mirror at negative distances (i.e.~the light first passes through the virtual mirror and then the atoms), and still have the m-CBS effect.
Fig.~\ref{fig:DistanceMirror} shows the measured dependence of the fringes' period $\Theta_f$ (or, equivalently, of the deduced mirror distance $\pi/k\theta_0\Theta_f$) as a function of $h$. The excellent linear behavior not only shows a good agreement with theory, but also indicates that the initial experimental positioning was misaligned by $2.1(1)~$mm.

\section{Saturated regime}

\begin{figure}
\includegraphics[width=7.8 truecm]{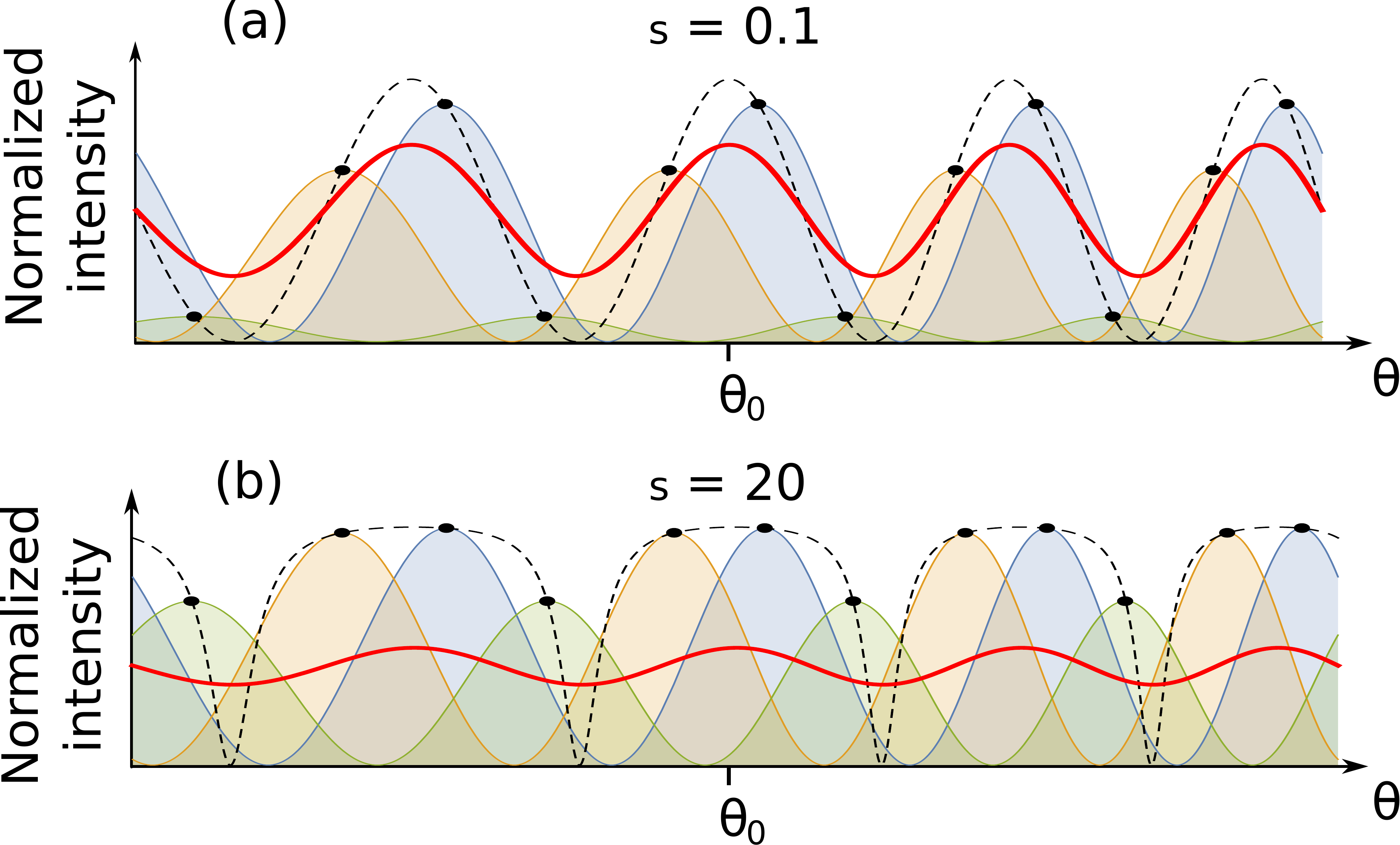}
	\caption{{Single-atom (colored areas) and cloud-averaged (red line) fringe patterns in the linear (a) and saturated (b) regime. The reason for the periodic modulation of the fringes' maxima for different atoms is explained in the main text; the black dashed line denotes the maximum amplitude of individual atomic fluorescence fringes, as a function of the maxima position. The maxima of all single-atom curves are identified by a black dot. In the saturated regime, the saturation of the atoms reduces the amplitude modulation, and the contrast of the averaged fringes decrease.}}
	\label{fig:Fringes}
\end{figure}

In the saturated regime, the atomic dipole moment has a nonlinear response to the applied radiation field. It is then no longer possible to add up linearly the amplitudes of four independent processes. In order to better understand how the contrast of m-CBS depends on the saturation parameter, we turn to an alternative picture, valid for all saturation, including the low and large saturation limits. In this new picture, we first consider the total fluorescence of a single atom and its mirror image. 
This fluorescence is a coherent superposition of the light scattered by the atom and its mirror image, which have same amplitude and are strongly correlated. 
This leads, for a single atom, to an undamped far field fringe pattern with full contrast. Furthermore, different atoms located at different distances from the virtual mirror are exposed to different local amplitudes of the standing wave created by the superposition of the incoming and reflected m-CBS beams (see Eq.~\ref{StandingWave} of Appendix~\ref{theory}). For a mirror of perfect reflectivity, and neglecting the attenuation of the m-CBS beam after its passage through the atomic cloud, both the incoming and reflected m-CBS beams have equal intensity, and the standing wave created has perfect nodes and antinodes. The absolute amplitude of the far field fluorescence fringes of a single atom is thus a function of the local light intensity at its position, which presents the non-homogeneous distribution of the standing wave.
Atoms at the maxima of the standing wave will thus have their far field fluorescence pattern with maximum amplitude (see blue line in Fig.~\ref{fig:Fringes}), compared to other atoms within the atomic cloud (e.g. yellow or green lines in Fig.~\ref{fig:Fringes}). Within the previous description based on four different processes, valid at low $s$, these atoms at the maxima of the standing wave are those that have all four processes summed up constructively, with a maximum of the scattered intensity at $\theta = \theta_0$. On the other hand, for different atomic positions in the standing wave, while the reciprocal processes (i) and (ii) have always constructive interference at $\theta = \theta_0$, the angular position of the maximum of the fluorescence for processes (iii) and (iv) varies and do not always cooperate to produce a maximum intensity at $\theta = \theta_0$. Hence, adding up coherently all four amplitudes for a single atom creates angular fringes which are a function of the atomic position, with different amplitudes, and maxima at different angles. An illustration of some of these single-atom angular fluorescence patterns are shown in Fig.~\ref{fig:Fringes} by the colored, filled curves. The black dashed curve shows the dependence of the amplitude of all possible single-atom fluorescence fringes as a function of the position of the maxima of the fringes. 
Considering an extended atomic sample leads to the superposition of shifted fringes with different amplitudes. Then, averaging over various disorder configurations leads to a fluorescence pattern with contrasted fringes around $\theta=\theta_0$ (shown by the thick red curve).

As expected, for low saturation ($s\ll 1$) this interpretation yields the same contrast $C=1$ as the interpretation based on reciprocal paths and an incoherent background. However this alternative interpretation of m-CBS allows us to go beyond the linear response theory and obtain quantitative predictions for the saturated m-CBS regime, which is illustrated by Fig.~\ref{fig:Fringes}(a) and Fig.~\ref{fig:Fringes}(b). For $s=20$, the position of the maxima of each individual atomic fringe is the same as for $s\ll 1$. Except for atoms in an ever narrower slice around the nodes of the standing wave produced by the incident laser, though, the amplitude of these individual fringes all saturate now to the same value. After averaging over all atoms, the contrast of the total detected intensity pattern, as illustrated by the red lines in Fig.~\ref{fig:Fringes}(b), is reduced.
Note that for the sake of simplicity, the calculation shown in Fig.~\ref{fig:Fringes} is done for atoms distributed over a small region of size of the order of a few wavelengths, and the envelope of the fringes (as expressed by the exponential function in Eq.\eqref{eq:mod}) is thus not visible.

\begin{figure}
	\includegraphics[width=.5\textwidth]{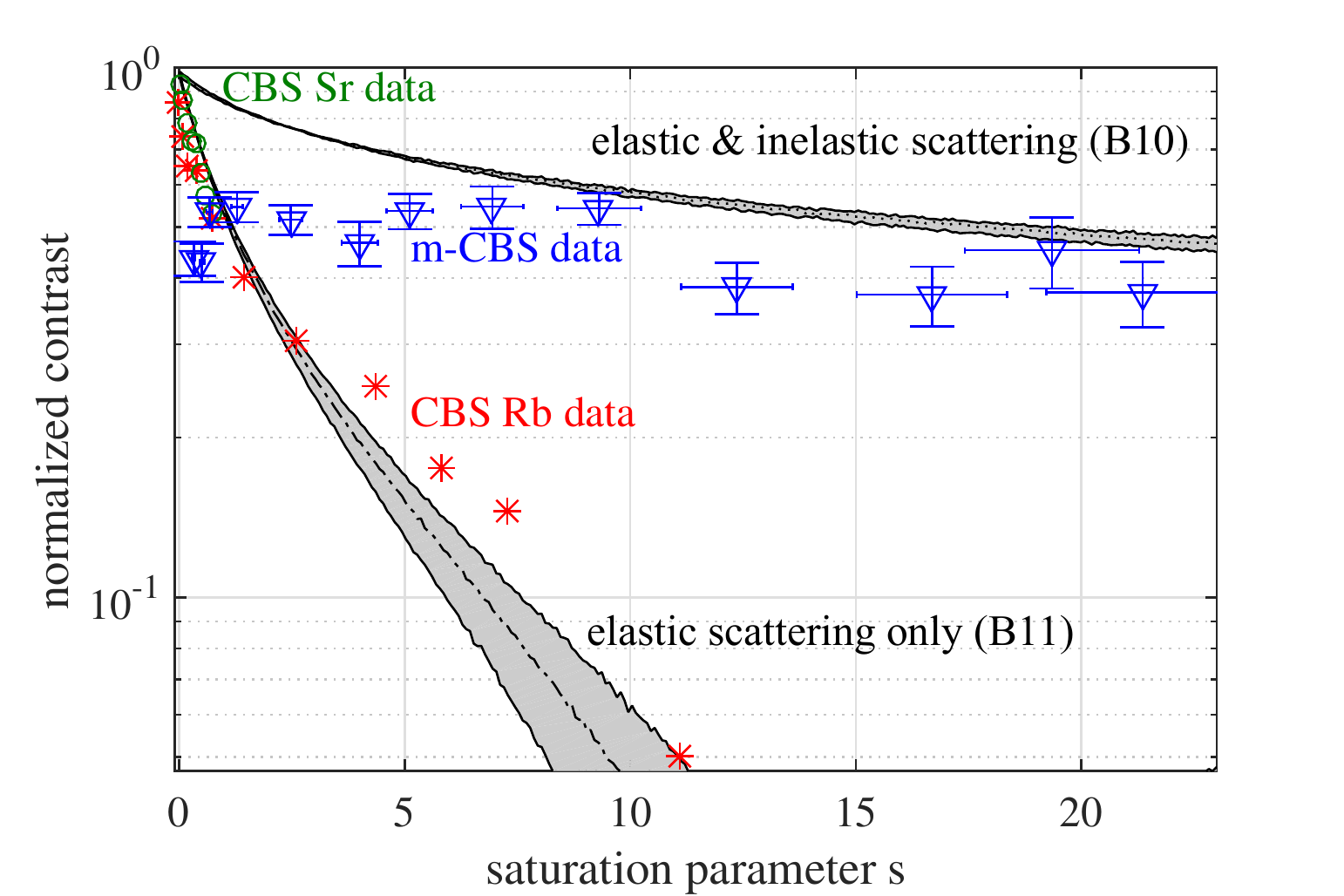}
	\caption{Experimental m-CBS contrast (blue triangles), compared to the normalized CBS one from a strontium cloud~\cite{Chaneliere04} (green circles) and from a rubidium cloud~\cite{Labeyrie08} (red stars). The gray areas for the theoretical predictions (for the elastic and inelastic theories (see main text and Appendix~\ref{theory})) account for the precision within which our experimental parameters are known: $h=8\pm0.5$mm and $\sigma_z=0.9\pm0.1$mm. The data are presented as a function of the saturation parameter $s$ at the center of the probe beam.}
	\label{fig:ContrastSaturation}
\end{figure}

In the discussion of the reduced contrast of CBS, another important argument has been the role of the inelastically scattered light, also known as the Mollow triplet. In the single scattering regime considered here, one can solve the optical Bloch equations of independent atoms, and then sum their (independent) scattered field. For each atom, one computes the field scattered by the oscillating dipole as being proportional to the optical coherence. In the limit of vanishing saturation, this allows to compute the total scattered intensity, as all light is elastically scattered. For larger saturation however, the optical coherence saturates and even decreases to zero for very large values of $s$. In contrast, the excited state population saturates to a non zero value and allows to compute the total scattered intensity. The difference between the light scattering computed from the atomic coherence or the excited state population corresponds to inelastically scattered light (sometimes interpreted as being stimulated by vacuum fluctuations).
For atoms separated by more than one wavelength, emission of such inelastic photons corresponds to randomly oscillating dipoles and are thus assumed not to preserve complete phase coherence with the incident laser, resulting in a  reduction of the CBS contrast, in addition to the nonlinear response. In m-CBS however, the two atoms contributing to the fringes (i.e. the atom and its mirror image) have strongly correlated oscillations, which preserve the relative phase, even when randomly oscillating. Thus, inelastic scattering is expected to fully contribute to the m-CBS effect, leading to a lower reduction of the contrast for increasing saturation, as compared to CBS. As detailed in the Appendix~\ref{theory} we have derived a prediction for the m-CBS contrast based on the elastically scattered light alone or based on the light scattered using the excited state population. The resulting predictions are indicated in Fig.~\ref{fig:ContrastSaturation} by "elastic scattering only" and "elastic and inelastic scattering", respectively. We note that, for CBS, self-interference of inelastic photons has been identified to lead to a finite contrast even in the very large saturation limit~\cite{Shatokhin05}. We also note that another different, hitherto unexplained scaling for CBS with rubidium has been reported in ~\cite{Kupriyanov06} and might be due to a modification of the internal states of the rubidium atoms~\cite{Balik05}. In any case, our model including elastic and inelastic scattering predicts m-CBS to be much more robust against saturation than CBS.

A precise measurement of the contrast and its dependence on the saturation parameter requires a careful data treatment, in order to fully eliminate the undesired additional signal at the camera created by the reflected m-CBS beam scattered from all the optical elements. The procedure for extracting the fluorescence signal out of our raw data is detailed in the Appendix~\ref{DataT}. For a given mirror position ($h = 8.0(5)~$mm) and cloud size ($\sigma_z = 0.9(1)~$mm), we have repeated the experimental sequence described before for several different intensities of the m-CBS beam. This allowed us to obtain a series of curves from which the absolute fringe contrast could be extracted. Fig.~\ref{fig:ContrastSaturation} confronts the saturation behavior of the enhancement factor measured in our m-CBS setup and conventional CBS measured by other groups, as a function of the saturation parameter $s$ at the center of the m-CBS beam. Note that we used the same definition for the contrast of CBS as we did for m-CBS before. The CBS data with strontium atoms from  ~\cite{Chaneliere04} have been obtained with a probe beam on resonance, and exhibit almost the maximum CBS contrast of $1$ in the low saturation regime. CBS was also reported with rubidium atoms which, due to their non-trivial internal structure, exhibit an overall lower contrast, even at low saturation parameter~\cite{Labeyrie99}. We have chosen, then, to present normalized values of the contrast for CBS rubidium data. Note that in order to observe CBS, a high optical thickness ($b>1$) is necessary to reach a significant contrast, whereas in the m-CBS set-up, the maximum contrast is achieved at low optical densities. 

\section{Discussion and conclusion}

The comparison between CBS and m-CBS data shows a completely different scaling for the contrast as a function of the saturation parameter. The m-CBS contrast follows the same scaling that we would expect from our model including elastic and inelastic scattering, allowing us to conclude that all photons contribute to m-CBS, whereas CBS interference is severely reduced in the presence of inelastic scattering. The remaining small discrepancy between theory and experiment might be explained by at least two different effects. The first one is the small, but non-zero, optical thickness of our cloud ($b\sim0.6$ at its center). At small saturation parameters, the partial re-absorption of the fluorescence light by other atoms from the cloud will change the relative amplitude of the paths of Fig.~\ref{fig:ScatScheme} that have otherwise equal amplitude, thus reducing the overall contrast. This is responsible for the large discrepancy between experimental data and the inelastic theoretical curve at small $s$. On the other hand, at high saturation parameters the spectral broadening of the atomic fluorescence (known as the Mollow triplet) may also reduce the m-CBS contrast, which is not accounted for in our model. The spectrum of inelastically scattered photons from a resonant probe beam presents sidebands displaced by $\pm \Omega_0$ from the atomic resonance \cite{Mollow69}. Since the optical path for the light scattered by the atoms is different from the optical path for the light scattered by their mirror images by twice the distance from the atomic cloud to the real mirror ($\Delta l \approx 1.2~$m in our case), the inelastic broadening will cause a relative dephasing between both when the spectral width becomes comparable to, or bigger than, $(2\pi)c/\Delta l \approx (2\pi)250~$MHz. For our highest experimental saturation parameter, $s = 20$, the separation in frequency between the Mollow sidebands is equal to $2 \Omega_0 = 2\sqrt{s}\,\Gamma \sim (2\pi)273~$MHz, so the loss of optical coherence in the interference process is already expected to affect the m-CBS contrast. The role of the inelastic spectral broadening in the interference of the light emitted by the atoms and their mirror images will be studied in a future work.

In conclusion, we have observed coherent backscattering from a cloud of cold atoms and its mirror image. Investigating the saturated regime allowed us to identify the important contribution of inelastic photons to the interference process, at odds with CBS. Because this coherence effect appears in completely different regimes as compared to CBS, such as low optical densities and high saturation parameters, it can represent a very important tool for probing coherences in the atomic scattered light where CBS is not anymore observable. In particular, by an appropriate use of waveplates and different polarization channels, the m-CBS set-up should allow to select specific interference paths, which is not possible for CBS.
More generally, the atom and its mirror image are strongly correlated, which allows this situation to probe non-classical light effects~\cite{Steiner95,Eschner01, Wilson03, Hetet11, Hoi15}.

\begin{acknowledgments}
We appreciate helpful insights from Thibaut Jonckheere and Dominique Delande. We thank William Gu\'erin for his critical reading of the manuscript. This work has received financial support from FAPESP, CNPq (PVE 400228/2014-9) and CAPES Brazilian agencies. 
\end{acknowledgments}

\appendix

\section{Data analysis: subtraction of the laser light}
\label{DataT}
 
\begin{figure}
	\includegraphics[width=8.5 truecm, height=4 truecm]{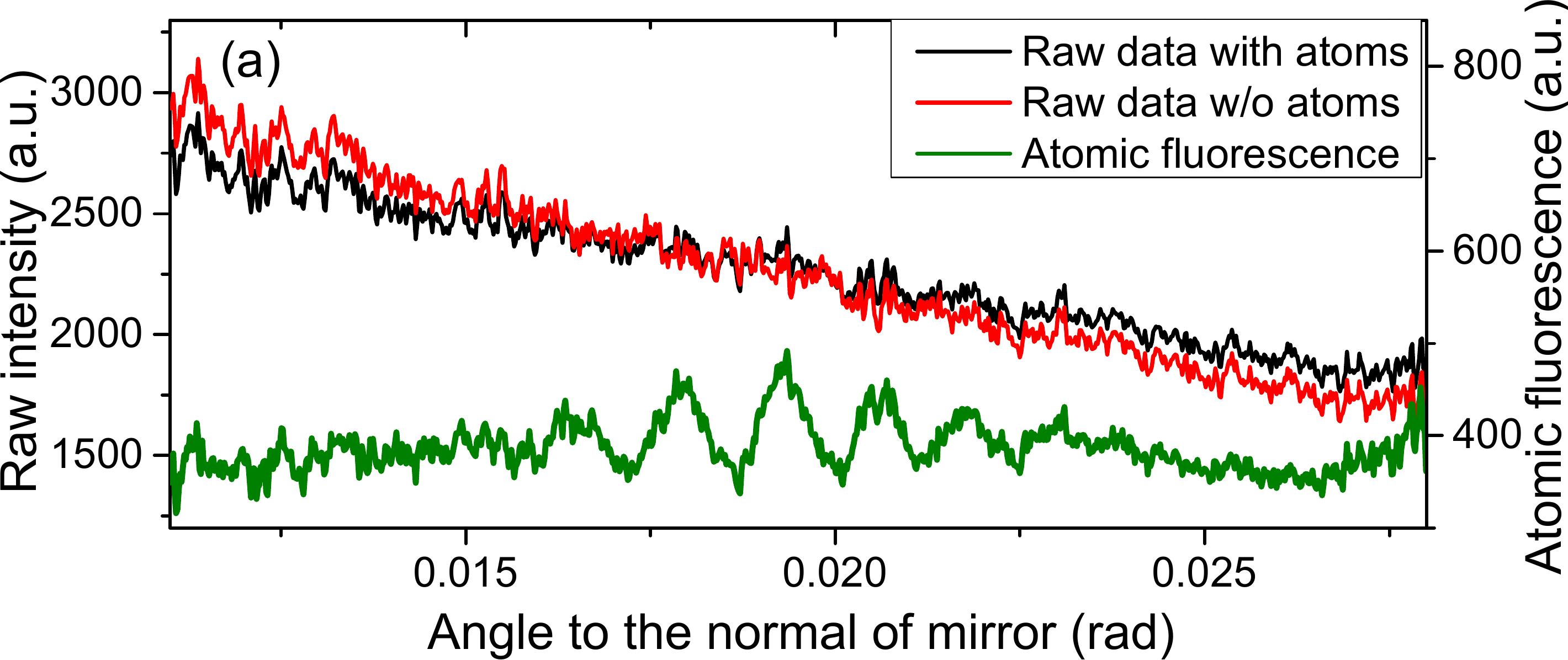}
	\\ \includegraphics[width=4.2 truecm]{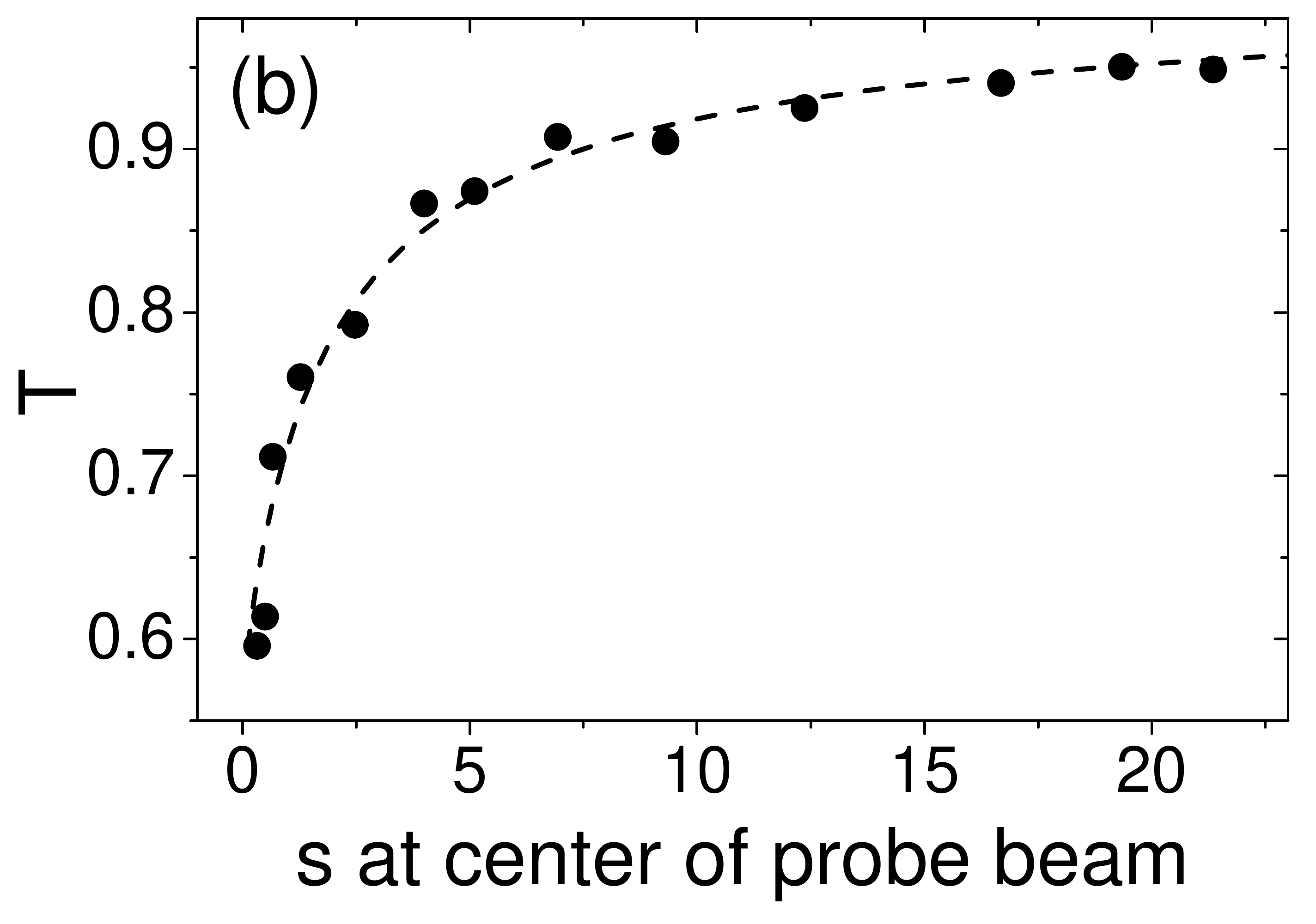}
	\hfil
	\includegraphics[width=4.2 truecm]{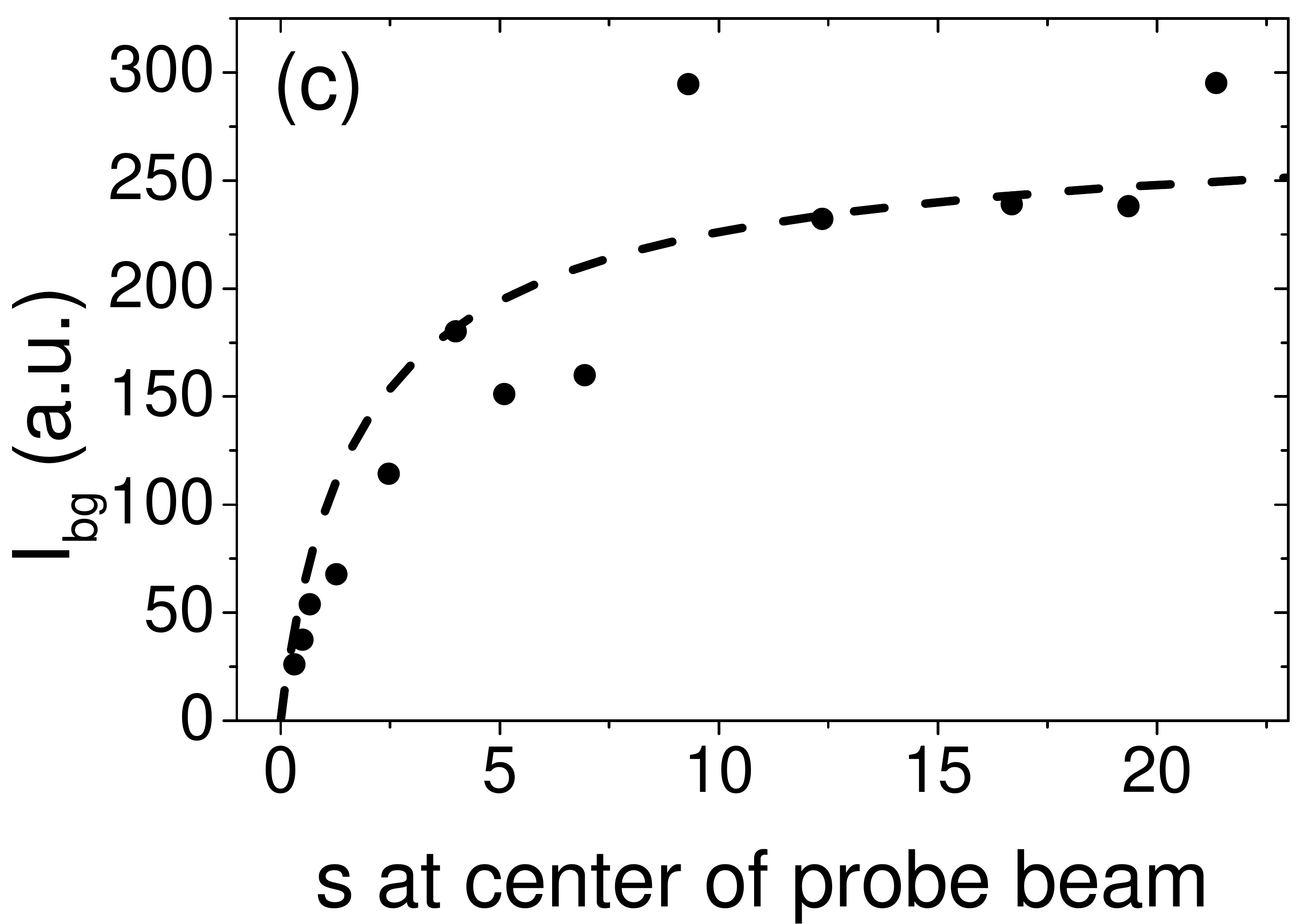}
	\caption{(a) Angular intensity profile in the presence (middle black curve) and in the absence (top red curve) of the atomic cloud. The deduced atomic fluorescence after the laser light weighted subtraction (see text) $I_f(\theta)$ (lower green curve, right vertical axis) presents a locally isotropic background, plus an interference pattern at the center. (b) Transmission coefficient $T$ and (c) background intensity $I_{\text{fluo}}$ deduced from the fit (see Eq.~\eqref{eq:stray2} and main text), as a function of the on-axis saturation parameter $s$. The dashed curves are calculated with a simple model for the interaction between our saturated Gaussian laser beam and our Gaussian atomic cloud.}
\label{fig:StrayLight}
\end{figure}

As described in the main text, at each experimental run we obtain two images: The first one is registered while the m-CBS beam impinges on the atomic cloud, and the second one is done in the same conditions, but with no atoms captured in our trap. The m-CBS reflected beam, after traversing all optical elements, creates an angular light profile on directions close to its propagation direction. The image registered in absence of atoms shows exclusively this light pattern; The image with atoms has this pattern superposed to the atomic fluorescence, that we want to extract. All the difficulty in extracting it stems from the fact that when the atoms are also present, the m-CBS beam is partially absorbed by them, which results in a smaller signal on the camera when compared to its effect without absorption. We can thus write the azimuthal-averaged profile of the light intensity in the presence of atoms as
\begin{equation}
I_a(\theta)=TI_{\text{las}}(\theta)+I_{f}(\theta),\label{eq:stray}
\end{equation}
where $T=e^{-2b}$ stands for the average transmission of the cloud after the double passage of the reflected m-CBS beam, and $I_{\text{las}}$ and $I_{f}$ for, respectively, the intensity of the laser light in the absence of atoms, and the intensity of the atomic fluorescence only. From $I_{f} (\theta)$, the absolute contrast can thus be extracted. Out of the fringes' envelope (or, to a good approximation, for $|\theta-\theta_0|>2\Phi$), and since we are monitoring the intensity in a narrow angle of $10~$mrad, the fluorescence background is isotropic to an excellent approximation. We can write then 
\begin{equation}
I_a(\theta,~|\theta-\theta_0|>2\Phi)=TI_{\text{las}}(\theta)+I_{\text{fluo}},\label{eq:stray2}
\end{equation}
where $I_{\text{fluo}}$ is the constant incoherent atomic background. Then, the laser light profile is determined by finding the linear combination of $I_a(\theta)$ (measured with the atoms) and $I_{\text{las}}(\theta)$ (measured without the atoms) that satisfies Eq.~\eqref{eq:stray2} outside of the fringes region. An example is provided in Fig.~\ref{fig:StrayLight}(a), where the measured intensity profile with and without the atoms is presented, as well as the extracted atomic fluorescence. Figs.~\ref{fig:StrayLight}(b) and (c) show the fitted parameters $T$ and $I_{\text{fluo}}$, respectively, as a function of the saturation parameter $s$ at the center of the atomic beam (black circles). The dashed lines correspond to a simple model for the interaction between our saturated Gaussian laser beam and our Gaussian shaped atomic cloud, with the optical density at the center as the only free parameter. The good agreement gives us confidence in the fitting procedure for extracting the pure atomic fluorescence out of our raw data.

\section{Single-scattering theory}
\label{theory}

In the following we outline the theoretical approach used to obtain the radiated intensity pattern in the linear and saturated regimes. Since we are focusing on the single scattering regime, it is sufficient to study the behavior of single atoms, and then sum their radiation independently. Let us thus consider a two-level atom at position $\mathbf{r}=\left(x,y,z\right)$  driven by a field of wavevector $\mathbf{k}_{0} = k\left(0,\sin\theta_{0},-\cos\theta_{0}\right)$.

Without loss of generality, we assume that the (virtual) dielectric mirror lies in the plane $z=0$ and choose the initial polarization along the $\hat{\mathbf{x}}$ axis. The incident field plus its mirror generate a standing wave in the $z$-axis and a propagating one in the $y$ direction:
\begin{equation}
\Omega\left(\mathbf{r}\right)=\Omega_{0}\cos (k z\cos\theta_0)e^{-i ky\sin\theta_0}.
\label{StandingWave}
\end{equation}
Note that we have here assumed high quality mirrors that possess unity reflection coefficients.

The single-atom equations are then used to determine the radiation of each atom. Calling $ \sigma$, $\sigma^{\dagger}$ and $\sigma^z$ the atomic operators, in the semiclassical limit the atomic dynamics is described by the following set of equations~\cite{Scully97}:
\begin{eqnarray}
\frac{d\hat\sigma }{dt}&=&\left(i\Delta-\frac{\Gamma}{2}\right) \hat\sigma +i\Omega\left(\mathbf{r}\right) \hat\sigma^{z} ,
\\ \frac{d\hat\sigma^{z}}{dt}&=&2i\left(\Omega^{*}\left(\mathbf{r}\right) \hat\sigma -\Omega\left(\mathbf{r}\right) \hat\sigma^{\dagger} \right)-\Gamma\left(\hat\sigma^{z}+1\right),
\end{eqnarray}
with the commutation relations $\left[\hat{\sigma},\hat{\sigma}^{z}\right]=2\hat{\sigma}$, $\left[\hat{\sigma}^{\dagger},\hat{\sigma}\right]=\hat{\sigma}^{z}$ and $\hat\sigma^{z}\hat{\sigma}=-\hat{\sigma}$.

By imposing the time derivatives to be zero, one obtains the steady state ($ss$) expectation values of the optical coherence and excited population for an atom at position $\mathbf{r}$:
\begin{eqnarray}
\left\langle \hat\sigma\right\rangle _{ss}(\mathbf{r})&=&\frac{\Delta^{2}+\Gamma^{2}/4}{\Delta+i\Gamma/2}\frac{\Omega(\mathbf{r})}{\Delta^{2}+\Gamma^{2}/4+2|\Omega(\mathbf{r})|^{2}},
\\ \left\langle \hat{\sigma}^{\dagger}\hat{\sigma}\right\rangle _{ss}(\mathbf{r})&=&\frac{|\Omega(\mathbf{r})|^2}{\Delta^{2}+\Gamma^{2}/4+2|\Omega(\mathbf{r})|^{2}}.
\end{eqnarray}

In the far field limit, the field radiated by a single atom in a direction $\mathbf{k}=k\hat{\mathbf{n}}$ and at a distance $r$ reads~\cite{Bienaime11}
\begin{align}
\hat{\mathbf{E}}\left(\mathbf{k},t\right) & =-\hat\sigma\left(t\right)\frac{d k^{2}}{4\pi\epsilon_{0}r}\left[\hat{\mathbf{\mathbf{n}}}\left(\hat{\mathbf{x}}\cdot\hat{\mathbf{n}}\right)-\hat{\mathbf{x}}\right]e^{-i\mathbf{k}\cdot\mathbf{r}},
\end{align}
where $d$ refers to the dipole coupling element, $\epsilon_0$ to the vacuum permittivity.

The measured field in the m-CBS experiment actually contains two contributions from each atom, since the radiation of the latter may be reflected or not by the mirror (see Fig.~\ref{fig:ScatScheme}, with processes (i) and (iv) that yield mirror reflection after scattering). Thus the total scattered electric field $\mathbf{E}_{s}$ in a direction $\mathbf{k}$ comes from the superposition
\begin{equation}
\mathbf{E}_{s}\left(\mathbf{k}\right)=\mathbf{E}\left(k_{x},k_{y},k_{z}\right)+\mathbf{E}\left(k_{x},k_{y},-k_{z}\right).
\end{equation}
In general, the different components of the field may play an important role in the 
intensity profile. However our experiment was carried out within observation angles $\theta\ll1$, where only the $E_x$ component is significant. Hence we obtain for the steady-state fields the following expression:
\begin{eqnarray}
\langle E_{s}\rangle&\sim&\sqrt{\alpha}\langle\sigma\rangle_{ss}\cos \left(kz\cos\theta\right)e^{-ik_{y}y-ik_{x}x},\label{eq:Es}
\\ \langle E_{s}^\dagger E_{s}\rangle&\sim&\alpha\left\langle \sigma^{\dagger}\sigma\right\rangle_{ss}\cos^{2}\left(kz\cos\theta\right),\label{eq:EsEs}
\end{eqnarray}
where the prefactor $\alpha = d^{2}k^{4}/(4\pi^{2}\epsilon_{0}^{2}r^{2})$ is unimportant to the determination of the contrast. Eq.~\eqref{eq:Es} corresponds to the optical coherence, and thus to the {\it elastically scattered} light. On the contrary Eq.~\eqref{eq:EsEs} describes the {\it total} intensity, that is, both elastic and inelastic photons~\cite{Scully97}.

Moving to the m-CBS by a macroscopic cloud of $N$ atoms with positions $\mathbf{r}_j$, the radiation pattern is computed as the sum of the single-atom intensities:
\begin{equation}
\frac{I_{\text{tot}}}{I_{0}} =\frac{4s}{N}\sum_{j=1}^{N}\frac{\cos^{2}\left(kz_{j}\cos\theta_{0}\right)\cos^{2}\left(kz_{j}\cos\theta\right)}{1+s\cos^{2}\left(kz_{j}\cos\theta_{0}\right)},\label{eq:Itot}
\end{equation}
where we have introduced the saturation parameter $s=2\Omega_{0}^{2}/(\Delta^{2}+\Gamma^{2}/4)$ and $I_{0}=c\epsilon_0N\alpha/8$. Eq.~\eqref{eq:Itot} relies on the simplifying hypothesis that all the scattered light (elastic {\it and} inelastic) has the same phase after the scattering, at any time. More precisely, the coherence length of the light is much larger than the distance between the (real) mirror and the cloud, so the inelasticity of the photons may not play a role. We note that, in the case of a uniform intensity distribution of the laser beam, the spatial distribution of the atoms in the plane of the mirror $(x,y)$ does not play any role
in this set-up.

Nevertheless, if one was to assume that only elastically scattered light contributes to the fringes pattern, the following expression for the intensity would be obtained:

\begin{equation}
\frac{I_{\text{ela}}}{I_{0}}=\frac{4s}{N}\sum_{j=1}^{N}\frac{\cos^{2}\left(kz_{j}\cos\theta_{0}\right)\cos^{2}\left(kz_{j}\cos\theta\right)}{\left(1+s\cos^{2}\left(kz_j\cos\theta_{0}\right)\right)^{2}},\label{eq:Iela}
\end{equation}

\begin{figure}	
	\hfil	\includegraphics[width=3.9 truecm]{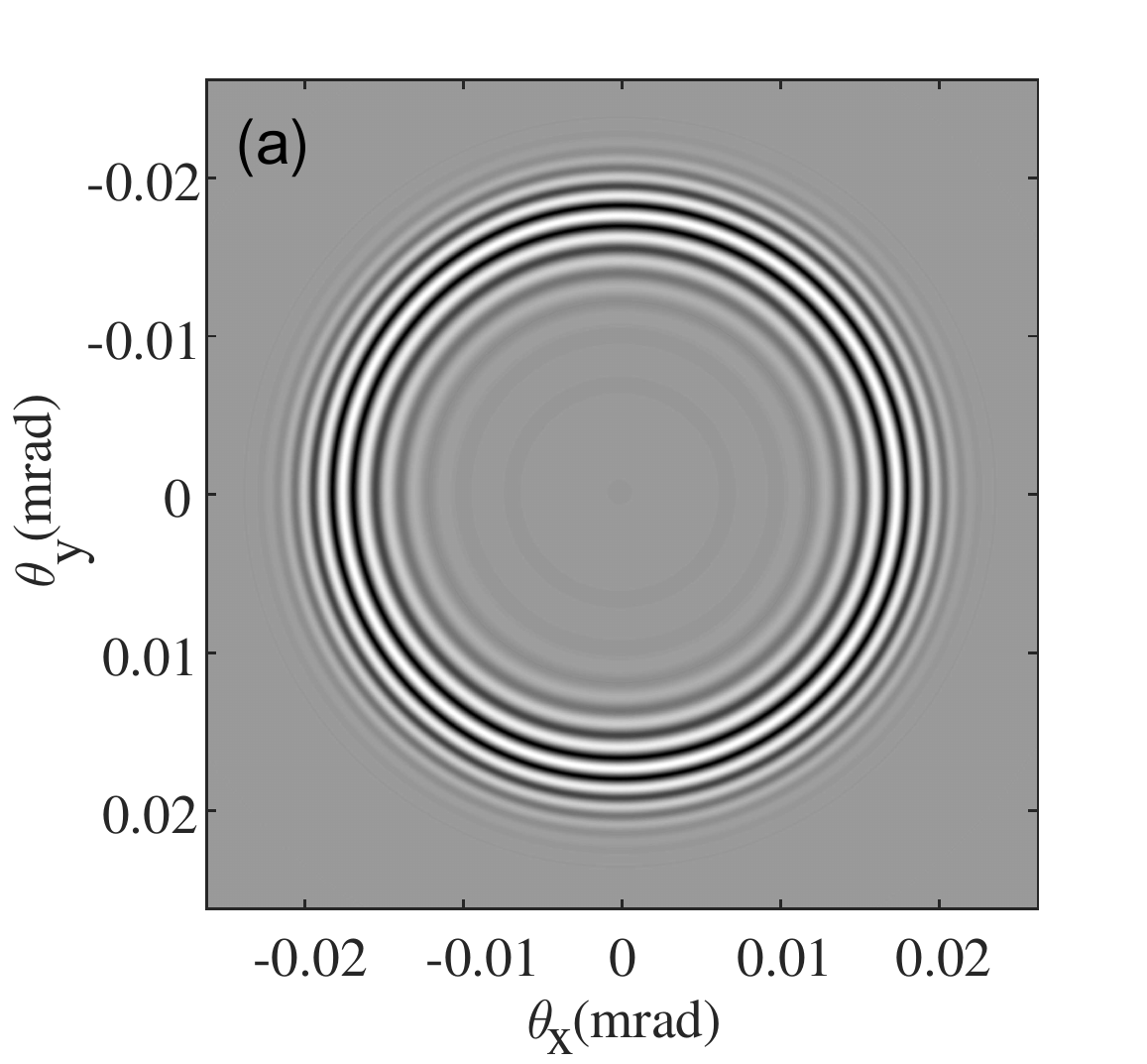}
 	\hfil \includegraphics[width=4.5 truecm]{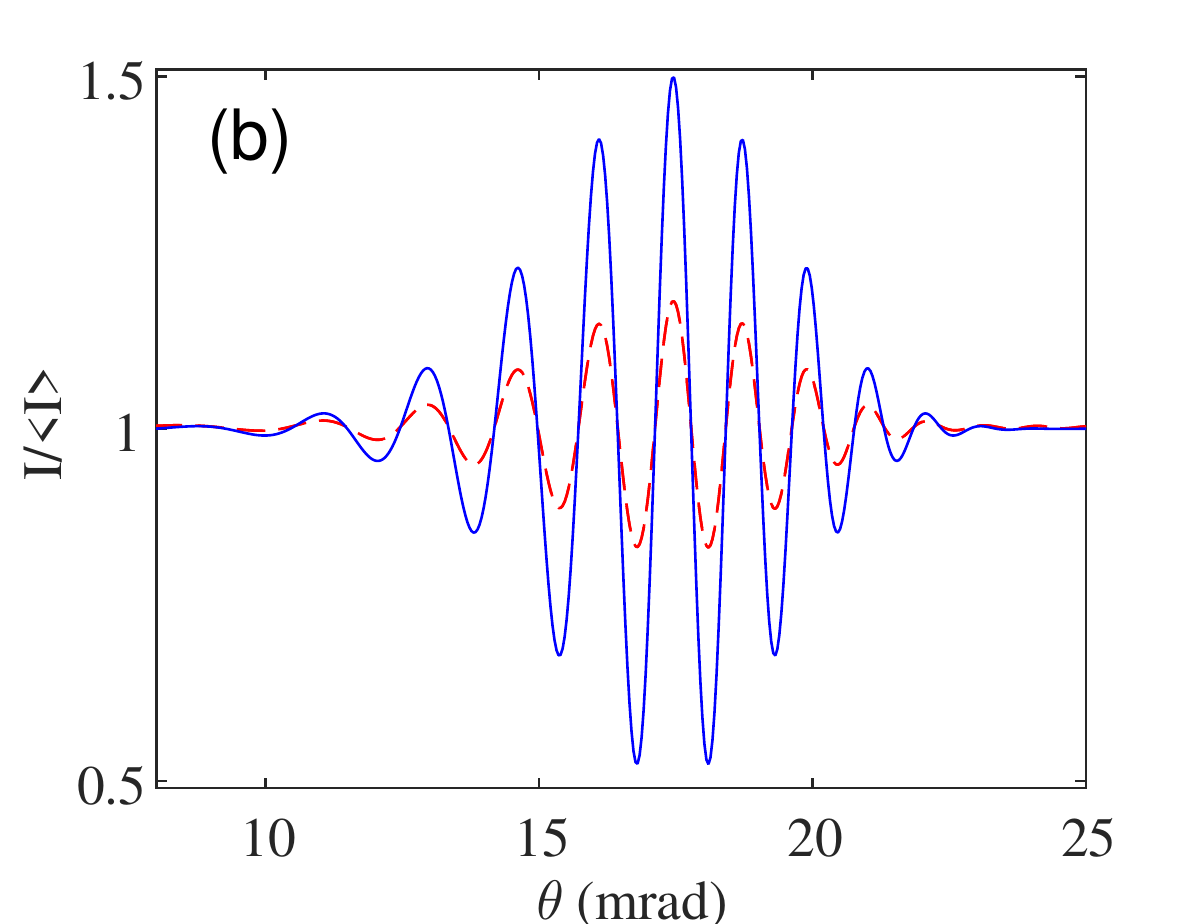}
	\caption{(a) Intensity pattern in the nonlinear regime (Eq.~\eqref{eq:Itot}), with $s = 20$. The intensity pattern in the linear regime is virtually identical, up to a scaling factor. (b) Azimuthal average of the intensity pattern normalized to the background intensity, in the linear ($s = 0.01$, blue continuous line) and nonlinear ($s = 20$, red dashed line) regimes, showing the fringes profile. The figures can be compared to the measurement presented in Fig.~2 of main text.}
	\label{fig:Theory}
\end{figure}

Let us first focus on the linear regime, which is that of elastic scattering ($I_{\text{tot}}\approx I_{\text{ela}}$), and that is obtained by taking $s\ll 1$. We obtain the following expression for the microscopic system:
\begin{eqnarray}
\frac{I_{\text{ela}}(\theta)}{I_0} &\approx \frac{s}{N}&\sum_{j=1}^{N}\left[ \cos\left(kz_{j}(\cos\theta-\cos\theta_{0})\right) \right. \nonumber
\\ &&\left.+\cos\left(kz_{j}(\cos\theta+\cos\theta_{0})\right)\right]^2,\label{eq:Iela1}
\end{eqnarray}
At this point, it must be noted that a single atom will exhibit full contrast, with fringes that are not damped. However, in the many body case, the superposition of atoms with different fringes phase (see Fig.~\ref{fig:Fringes}) results in an interference pattern with a reduced contrast, over a finite envelope.
This sum over a macroscopic cloud (i.e., much larger than the wavelength) is well captured by substituting the sum over the atoms by an integral over the typical atomic distribution $\sum_{j=1}^{N}\rightarrow\int\mbox{d}\mathbf{r}\rho(\mathbf{r})$. The first term in the sum in Eq.~\eqref{eq:Iela1} provides a {\it coherent} contribution in the $\theta=\theta_0$ direction, whatever the position $z$ of the atom, whereas the second term averages to $0$. The decay of the envelope is then provided by the finite size of the cloud.
For a Gaussian distribution such as those produced in our trap, one obtains Eq.~1. That equation predicts an alternation of constructive and destructive interferences with period $\pi/(kh\theta_0)$, which leads to the observed fringes, see Fig.~\ref{fig:Theory}.

Remark that the elastic contribution Eq.~\eqref{eq:Iela} decreases as $1/s$ for increasing saturation parameter $s$, whereas the total radiation converges to:
\begin{equation}
\lim_{s\rightarrow\infty}\frac{I_{\text{tot}}(\theta)}{I_{0}}=\frac{4}{N}\sum_{j=1}^{N}\cos^{2}\left(kz_{j}\cos\theta\right).\label{sinf}
\end{equation}
Integrating over a Gaussian distribution as before, in the small angle and large cloud limits, we obtain $I_{\text{tot}}=2I_0$, i.e., the fluorescence converges to a finite value for very large saturation parameters. Thus the ratio between the elastically scattered intensity to the total one scales as $1/s$, which explains the fast decay of the contrast of the 'elastic' theory.

In the saturated regime, atoms that are not close to a zero of the standing wave saturate. As $s$ increases, the proportion of scatterers that contribute to the grating decreases as $1/\sqrt{s}$, whereas the others produce an isotropic fluorescence radiation pattern. This explains the rather slow decay of the contrast in the 'inelastic theory'. In the present work, the intensity pattern for large values of $s$ was computed numerically using the microscopic formula \eqref{eq:Itot}, for random Gaussian distributions of millions of atoms (see Fig.~\ref{fig:ContrastSaturation}).

\end{document}